# New method for the extrapolation of finite-size data to infinite volume

Sergio Caracciolo[a], Robert G. Edwards[b], Sabino J. Ferreira[c], Andrea Pelissetto[d] and Alan D. Sokal[e]*

[a]Dipartimento di Fisica and INFN, Università degli Studi di Lecce, Lecce 73100, ITALIA

[b]SCRI, Florida State University, Tallahassee, FL 32306, USA

[c]Departamento de Física, Universidade Federal de Minas Gerais, Belo Horizonte, MG 30161, BRASIL

[d]Dipartimento di Fisica and INFN – Sezione di Pisa, Università degli Studi di Pisa, Pisa 56100, ITALIA

[e]Department of Physics, New York University, 4 Washington Place, New York, NY 10003, USA

We present a simple and powerful method for extrapolating finite-volume Monte Carlo data to infinite volume, based on finite-size-scaling theory. We discuss carefully its systematic and statistical errors, and we illustrate it using three examples: the two-dimensional three-state Potts antiferromagnet on the square lattice, and the two-dimensional $O(3)$ and $O(\infty)$ $\sigma$-models. In favorable cases it is possible to obtain reliable extrapolations (errors of a few percent) even when the correlation length is 1000 times larger than the lattice.

Quantum field theorists are interested primarily in infinite systems; but Monte Carlo simulations must perforce be carried out on lattices of finite linear size $L$, limited by computer memory and speed. This raises the problem of extrapolating finite-volume data to $L = \infty$. We present here a simple and powerful method for performing this extrapolation, based on finite-size-scaling theory [1]; and we discuss carefully its systematic and statistical errors. We illustrate the method using three examples: the two-dimensional three-state Potts antiferromagnet on the square lattice [2], and the two-dimensional $O(3)$ and $O(\infty)$ $\sigma$-models [3,4]. We have found — much to our surprise — that in favorable cases it is possible to obtain reliable extrapolations (errors of a few percent) at $\xi/L$ as large as 10–1000. More details can be found in [5].

Consider, for starters, a model controlled by a renormalization-group (RG) fixed point having *one* relevant operator. Let us work on a periodic lattice of linear size $L$. Let $\xi(\beta, L)$ be a suitably defined finite-volume correlation length (we use the second-moment correlation length defined by equations (4.11)–(4.13) of [6]), and let $\mathcal{O}$ be any long-distance observable (e.g. the correlation length or the susceptibility). Then finite-size-scaling theory [1] predicts that

$$\frac{\mathcal{O}(\beta, L)}{\mathcal{O}(\beta, \infty)} = f_{\mathcal{O}}\!\left(\xi(\beta, \infty)/L\right) + O\!\left(\xi^{-\omega}, L^{-\omega}\right) \quad (1)$$

where $f_{\mathcal{O}}$ is a universal function and $\omega$ is a correction-to-scaling exponent. Hence, if $s$ is any fixed scale factor (usually we take $s = 2$),

$$\frac{\mathcal{O}(\beta, sL)}{\mathcal{O}(\beta, L)} = F_{\mathcal{O}}\!\left(\xi(\beta, L)/L\right) + O\!\left(\xi^{-\omega}, L^{-\omega}\right) \quad (2)$$

where $F_{\mathcal{O}}$ can be expressed in terms of $f_{\mathcal{O}}, f_{\xi}$.

Our method proceeds as follows [7]: Make Monte Carlo runs at numerous pairs $(\beta, L)$ and $(\beta, sL)$. Plot $\mathcal{O}(\beta, sL)/\mathcal{O}(\beta, L)$ versus $\xi(\beta, L)/L$, using those points satisfying both $\xi(\beta, L) \geq$ some value $\xi_{min}$ and $L \geq$ some value $L_{min}$. If all these points fall with good accuracy on a single curve — thus verifying the Ansatz (2) for $\xi \geq \xi_{min}, L \geq L_{min}$ — choose a smooth fitting function $F_{\mathcal{O}}$. Then, using the functions $F_{\xi}$ and $F_{\mathcal{O}}$, extrapolate the pair $(\xi, \mathcal{O})$ successively from $L \to sL \to s^2 L \to \ldots \to \infty$.

We have chosen to use functions $F_{\mathcal{O}}$ of the form

$$F_{\mathcal{O}}(x) = 1 + a_1 e^{-1/x} + \ldots + a_n e^{-n/x} \quad (3)$$

This form is partially motivated by theory, which tells us that $F(x) \to 1$ exponentially fast as $x \to$

---

*Speaker at the conference.

0 [10]. Typically a fit of order $3 \leq n \leq 12$ is sufficient; we increase $n$ until the $\chi^2$ of the fit becomes essentially constant. The resulting $\chi^2$ value provides a check on the systematic errors arising from corrections to scaling and/or from the inadequacies of the form (3).

The *statistical* error on the extrapolated value of $\mathcal{O}_\infty(\beta) \equiv \mathcal{O}(\beta, \infty)$ comes from three sources: (i) error on $\mathcal{O}(\beta, L)$, which gets multiplicatively propagated to $\mathcal{O}_\infty$; (ii) error on $\xi(\beta, L)$, which affects the argument $x \equiv \xi(\beta, L)/L$ of the scaling functions $F_\xi$ and $F_\mathcal{O}$; and (iii) statistical error in our estimate of the coefficients $a_1, \ldots, a_n$ in $F_\xi$ and $F_\mathcal{O}$. The errors of type (i) and (ii) depend on the statistics available at the single point $(\beta, L)$, while the error of type (iii) depends on the statistics in the whole set of runs. Errors (i)+(ii) [resp. (i)+(ii)+(iii)] can be quantified by performing a Monte Carlo experiment in which the input data at $(\beta, L)$ [resp. the whole set of input data] are varied randomly within their error bars and then extrapolated.

The discrepancies between the extrapolated values from different lattice sizes at the same $\beta$ — to the extent that these exceed the estimated statistical errors — indicate the presence of systematic errors and thus the necessity of increasing $L_{min}$ and/or $\xi_{min}$ and/or $n$.

A figure of (de)merit of the method is the relative variance on the extrapolated value $\mathcal{O}_\infty(\beta)$, multiplied by the computer time needed to obtain it. We expect this *relative variance-time product* [for errors (i)+(ii) only] to scale as

$$\text{RVTP}(\beta, L) \approx \xi_\infty(\beta)^{d+z_{int,\mathcal{O}}} G_\mathcal{O}\Big(\xi_\infty(\beta)/L\Big) \quad (4)$$

where $d$ is the spatial dimension and $z_{int,\mathcal{O}}$ is the dynamic critical exponent of the Monte Carlo algorithm being used; here $G_\mathcal{O}$ is a combination of several static and dynamic finite-size-scaling functions, and depends both on the observable $\mathcal{O}$ and on the algorithm but not on the scale factor $s$. As $\xi_\infty/L$ tends to zero, we expect $G_\mathcal{O}$ to diverge as $(\xi_\infty/L)^{-d}$ (it is wasteful to use a lattice $L \gg \xi_\infty$). As $\xi_\infty/L$ tends to infinity, we expect $G_\mathcal{O} \sim (\xi_\infty/L)^p$ [5], but *the power $p$ can be either positive or negative*. If $p > 0$, there is an optimum value of $\xi_\infty/L$; this determines the best lattice size at which to perform runs for a given $\beta$. If $p < 0$, it is most efficient to use the *smallest* lattice size for which the corrections to scaling are negligible compared to the statistical errors.

Our first example [2] is the two-dimensional three-state Potts antiferromagnet on the square lattice, which is believed to have a critical point at $\beta = \infty$ [11]. We used the Wang-Swendsen-Kotecký cluster algorithm [12], which appears to have *no* critical slowing-down ($\tau_{int,\mathcal{M}^2_{stagg}} < 5$ uniformly in $\beta$ and $L$) [2]. We ran on lattices $L = 32, 64, 128, 256, 512, 1024, 1536$ at 153 different pairs $(\beta, L)$ in the range $5 \lesssim \xi_\infty \lesssim 20000$. Each run was between $2 \times 10^5$ and $2.2 \times 10^7$ iterations, and the total CPU time was modest by our standards (about 2 years on an IBM RS-6000/370). We took $\xi_{min} = 10$ and $L_{min} = 128$ and used a quintic fit in (3); the result for $F_\xi$ is shown in [2,5] ($\chi^2 = 75.41$, 66 DF, level = 20%). The extrapolated values from different lattice sizes at the same $\beta$ agree within the estimated statistical errors ($\chi^2 = 43.03$, 75 DF, level > 99%). The result for $G_\xi$ is shown in [5]: the errors are roughly constant for $\xi_\infty/L \gtrsim 0.4$ but rise sharply for smaller $\xi_\infty/L$. In practice we were able to obtain $\xi_\infty$ to an accuracy of about 1% (resp. 2%, 3%, 5%) at $\xi_\infty \approx 1000$ (resp. 2000, 5000, 10000).

Next let us consider [3,4] the two-dimensional $O(3)$ $\sigma$-model (see Caracciolo's talk for more details). We used the Wolff embedding algorithm with standard Swendsen-Wang updates; again critical slowing-down appears to be completely eliminated. We ran on lattices $L = 32, 48, 64, 96, 128, 192, 256, 384, 512$ at 180 different pairs $(\beta, L)$ in the range $20 \lesssim \xi_\infty \lesssim 10^5$. Each run was between $10^5$ and $5 \times 10^6$ iterations, and the total CPU time was 7 years on an IBM RS-6000/370. We took $\xi_{min} = 20$ and used a tenth-order fit. There appear to be weak corrections to scaling (of order $\lesssim 1.5\%$) in the region $0.3 \lesssim \xi_L/L \lesssim 0.7$ for lattices with $L \lesssim 64$–$96$. We therefore chose $L_{min} = 128$ for $\xi_L/L \leq 0.7$, and $L_{min} = 64$ for $\xi_L/L > 0.7$. The result for $F_\xi$ is shown in [4,5] ($\chi^2 = 72.91$, 73 DF, level = 48%). The result for $G_\xi$ is shown in [5]; at large $\xi_\infty/L$ it *decreases* sharply, with a power $p \approx -2$ in agree-ment with theory [5]. In practice we obtained $\xi_\infty$ to an accuracy of about 0.2% (resp. 0.7%, 1.1%, 1.6%) at $\xi_\infty \approx 10^2$ (resp. $10^3$, $10^4$, $10^5$).

We also carried out a "simulated Monte Carlo" experiment for the $O(N)$ $\sigma$-model at $N = \infty$, by generating data from the exact finite-volume solution plus random noise of 0.1% for $L = 64, 96, 128$, 0.2% for $L = 192, 256$ and 0.5% for $L = 384, 512$ [which is the order of magnitude we attain in practice for $O(3)$]. We considered 35 values of $\beta$ in the range $20 \lesssim \xi_\infty \lesssim 10^6$. We used $\xi_{min} = 20$ and $L_{min} = 64$ (in fact much smaller values could have been used, as corrections to scaling are here very small) and a ninth-order fit; for two different data sets we get $\chi^2 = 114$ (resp. 118) with 166 DF. In practice we obtain $\xi_\infty$ with an accuracy of 0.6% (resp. 1.2%, 2%, 3%) at $\xi_\infty \approx 10^3$ (resp. $10^4$, $10^5$, $10^6$). Here we can also compare the extrapolated values $\xi_\infty^{extr}(\beta)$ with the exact values $\xi_\infty^{exact}(\beta)$. Defining $\mathcal{R} = \sum_\beta [\xi_\infty^{extr}(\beta) - \xi_\infty^{exact}(\beta)]^2 / \sigma^2(\beta)$, we find for the two data sets $\mathcal{R} = 17.19$ (resp. 25.81) with 35 DF. Only 6 (resp. 9) points differ from the exact value more than one standard deviation, and none by more than two.

Details on all of these models will be reported separately [2,4].

The method is easily generalized to a model controlled by an RG fixed point having $k$ relevant operators. It suffices to choose $k - 1$ dimensionless ratios of long-distance observables, call them $R = (R_1, \ldots, R_{k-1})$; then the function $F_\mathcal{O}$ will depend parametrically on $R(\beta, L)$. In practice one can divide $R$-space into "slices" within which $F_\mathcal{O}$ is empirically constant within error bars, and perform the fit (3) within each slice. We have used this approach to study the mixed isovector/isotensor $\sigma$-model, taking $R$ to be the ratio of isovector to isotensor correlation length [3,4].

The method can also be applied to extrapolate the exponential correlation length (inverse mass gap). For this purpose one must work in a system of size $L^{d-1} \times T$ with $T \gg \xi_{exp}(\beta, L)$ (cf. [8]).

We wish to thank Martin Hasenbusch and especially Jae-Kwon Kim for sharing their data with us, and for challenging us to push to ever larger values of $\xi/L$. This research was supported by CNR, INFN, CNPq, FAPEMIG,

DOE contracts DE-FG05-85ER250000 and DE-FG05-92ER40742, NSF grant DMS-9200719, and NATO CRG 910251.